\documentclass[a4paper,11pt]{article}
\pdfoutput=1 

\usepackage{jcappub} 
\usepackage[T1]{fontenc} 
\usepackage{amsmath}
\usepackage{appendix}

\newcommand{\vect}[1]{\boldsymbol{#1}}

\newcommand{\absum}[0]{\textsc{AbacusSummit} }
\newcommand{\compaso}[0]{\textsc{CompaSO} }

\title{\boldmath Halo Occupation Distribution of Emission Line Galaxies: fitting method with Gaussian
Processes}

\author[a,1]{Antoine Rocher,\note{Corresponding author.}}
\author[a]{Vanina Ruhlmann-Kleider,}
\author[a]{Etienne Burtin,}
\author[a]{Arnaud de Mattia}

\affiliation[a]
{Universit\'e Paris-Saclay, CEA, Institut de recherche sur les lois Fondamentales de l'Univers, 91191, Gif-sur-Yvette, France.}

\emailAdd{antoine.rocher@cea.fr}
\emailAdd{vanina.ruhlmann-kleider@cea.fr}
\emailAdd{etienne.burtin@cea.fr}
\emailAdd{arnaud.de-mattia@cea.fr}

\abstract{The halo occupation distribution (HOD)
framework is an empirical method to describe the connection between dark matter halos and galaxies, which is constrained by small scale clustering data. Efficient fitting procedures are required to scan the HOD parameter space. This paper describes such a method based on Gaussian Processes to iteratively build a surrogate model of the posterior of the likelihood surface from a reasonable amount of likelihood computations, typically two orders of magnitude less than standard Monte Carlo Markov chain algorithms. Errors in the likelihood computation due to stochastic HOD modelling are also accounted for in the method we propose. We report results of reproducibility, accuracy and stability tests of the method derived from simulation, taking as a test case star-forming emission line galaxies, which constitute the main tracer of the Dark Energy Spectroscopic Instrument and have so far a poorly constrained galaxy-halo connection from observational data.
}

\begin{document}
\maketitle

\section{Introduction}
In recent years, clustering measurements from large spectroscopic surveys of galaxies and quasars have provided an important probe of cosmology, through the precise determination of the baryon acoustic oscillation (BAO) scale and the estimation of the linear growth rate of structure through redshift space distortions (RSD) measurements.

Clustering analyses are tested on simulated catalogues of galaxies in large interval of scales spanning the linear and quasi-linear regimes. These tests rely heavily on N-body simulations coupled to prescriptions to describe the connection between dark matter halos and galaxies. Among these, the halo occupation distribution (HOD) is an empirical approach whose parameters are derived from clustering measurements at small scales, in a range that overlaps minimally with scales used in the standard analyses for BAO and RSD measurements.

Several approaches have been proposed to perform HOD fits efficiently, like tabulated HOD methods that pre-compute halo and particle clustering and convolve it with halo occupation distribution~\cite{TabCorr} or the recent AbacusHOD~\cite{AbacusHOD} pipeline, that first initialises random numbers for every halo and particle, down-samples halos and particles from N-body simulations, and then run inference on HOD parameters from a sampler to derive best-fit parameters. 
Gaussian Processes (GP)~\cite{Rasmussen_Williams_2005} offer an alternative route to build a model function that fits a set of observational data. 
In cosmology, GP have been recently used to build emulators to perform cosmological inference, whether iterative~\cite{Pellejero-Ibanez20, Neveux21, GPry_2022} or not~\cite{Aemulus, Nishimichi19}. In particular, reference~\cite{GPry_2022} is very instructive on the challenges to be faced to set up such a methodology.
GP are also very efficient to perform global optimisation of expensive and stochastic functions~\cite{Garnett2008}, as often encountered in HOD modelling.

This paper describes a method to fit HOD parameters on small scale clustering measurements using Gaussian Processes to create a surrogate model of the HOD likelihood posterior surface that allows 
parameter marginalisation and confidence contour derivation to be performed efficiently and in a faster way than standard methods based on Markov chain Monte Carlo (MCMC) algorithms, despite uncertainties in the $\chi^2$ estimates due to stochastic HOD modelling and cosmic variance. This method is intended to be applied to early data of the Dark Energy Spectroscopic Instrument (DESI) whose main program of observations started in May, 2021. 

The galaxies considered in this work are Emission Line Galaxies (ELGs), the main target class of DESI which should collect 17 million ELG redshifts after five years of observations~\cite{DESI_2016}. In two months of early observations, DESI observed 267k ELGs that constitute the data sample for which the method hereafter described is primarily intended for, but it could easily be extended to other galaxy tracers. HOD fitting for ELGs is challenging since these galaxies are expected to reside in low mass halos $\sim10^{12} M_{\odot}$, which requires to handle high resolution N-body simulations where dark matter halos are well-resolved down to (at least) $10^{11} M_{\odot}$, making the evaluation of the HOD model, and the estimation of clustering statistics, CPU and memory expensive (due to the fast increase of the halo mass function at low mass). 

The outline of the paper is as follows. Section~\ref{sec:hod} introduces the ELG HOD modelling used in this paper. Section~\ref{sec:mock} describes the simulation and the clustering statistics. The 
inference methodoloy is detailed in section~\ref{sec:methodology}. Tests of the method are given in section~\ref{sec:tests}. We conclude in section~\ref{sec:conclusions}.

\section{HOD modelling}
\label{sec:hod}
The HOD formalism describes the relation between a typical class of galaxies and dark matter halos, as the probability that a halo with mass $M$ contains $N$ such galaxies. It also specifies how galaxy positions and velocities are distributed within halos. HOD models have contributions from two galaxy populations, namely centrals and satellites, with $\langle N_{cent}(M)\rangle$ and $\langle N_{sat}(M)\rangle$ their respective mean numbers hosted per halo of a given halo mass. 
Analytical descriptions of these mean HODs have been derived from either semi-analytical models or hydrodynamical simulations of galaxy formation and evolution (e.g.~\cite{Berlind03,Zheng05}). 

The most common mean HOD function
~\citep{Zheng07} uses a step function for centrals, a power law for satellites and assumes generally that satellites can only be found in halos which already host a central galaxy. This model has proven to describe well the clustering of luminosity selected~\cite{Zehavi11} or stellar mass limited~\cite{Contreras13} samples, like Luminous Red Galaxies (LRGs)~\cite{Zheng08}
or quasars~\cite{Smith20}.
However, a step function cannot represent 
young or star-forming galaxies~\citep{Zheng05,Contreras13}, which are  typical of ELG samples selected in recent surveys. 
As shown in~\cite{Contreras13,GonzalezPerez18}, the mean HOD predicted by ELG semi-analytical models can be fit reasonably well by a simple Gaussian or slightly asymmetric Gaussian for centrals, together with a power law for satellites. This was confirmed on data from the extended Baryon Oscillation Spectroscopic Survey~\cite{Avila20,Lin20}.

Accordingly, the baseline model we will follow in this paper, hereafter called Gaussian HOD model (GHOD), corresponds to:
\begin{equation}
\begin{split}
    & \left\langle N_{cent}(M)\right\rangle = \frac{A_c}{\sqrt{2\pi}\sigma_{\textsc{m}}} \cdot e^{-\frac{(\log_{10}(M)-\mu)^2}{2\sigma_{\textsc{m}}^2}}
        \\
    & \left\langle N_{sat}(M)\right\rangle = A_s\bigg(\frac{M-M_0}{M_1}\bigg)^{\alpha}
\end{split}
\label{GHOD}
\end{equation}

$A_c$ defines the amplitude of the central galaxy HOD, $\mu = \log_{10} M_c$ where $M_c$ is the characteristic halo mass 
with maximal probability to host a central galaxy and $\sigma_{\textsc{m}}$ is the 
width of the distribution. $A_s$ defines the amplitude of the satellite galaxy HOD, $M_{0}$ is a cut-off halo mass below which no satellite can be present, $\alpha$ is a slope-parameter that controls the variation in satellite richness with increasing halo mass, and $M_1$ is the mass at which 1 satellite is expected per halo if $A_s=1$ and 
$M_0$ is much lower than $M_1$.
 The total number density of the galaxy sample can be calculated as follows:

\begin{equation}
\bar{n}_{\mathrm{gal}} =\int \frac{\mathrm{d} n(M)}{\mathrm{d} M}\left[\left\langle N_{cent}(M)\right\rangle+\left\langle N_{sat}(M)\right\rangle\right] \mathrm{d} M
\label{ngal}
\end{equation}

The total galaxy sample size is governed by both $A_c$ and $A_s$  and the fraction of satellites is controlled by their ratio. Moreover, all other conditions being equal, the same clustering is obtained  whatever $A_c$ and $A_s$ values, provided their ratio is fixed and $\left\langle N_{cent}(M)\right\rangle$ remains lower than 1 (which is the case for all the fits performed in this paper).
We rely on this property to impose a fixed density in our fitting procedure.

The constraints adopted in our fits are the following.
First, as $\log_{10}(M_1)$ cannot be constrained due to degeneracies with $A_s$ and $\alpha$, this parameter is kept fixed throughout the paper. 
To choose a sensible value for $\log_{10}(M_1)$, we follow~\cite{Avila20} and set:
\begin{equation}
\log_{10}(M_1) = \log_{10}(M_c)^{ref} + 0.3 = 11.93
\end{equation}
taking for $\log_{10}(M_c)^{ref}$ the value used to generate our pseudo-data catalogues (see Table~\ref{tab:pseudo-data}).
Second, in order to apply a density constraint to our fits to match that in DESI data, we treat the $A_c$ and $A_s$ parameters in the following way. At each point in the HOD parameter space, $A_c$ is set to an initial value of 0.05, while $A_s$ is sampled from a flat prior range (reported in Table~\ref{tab:pseudo-data}). The total number density for these initial values of $A_c$ and $A_s$ is computed according to Eq.~\eqref{ngal} and the values of $A_c$ and $A_s$ are rescaled by the same factor (to preserve the clustering) in order to provide a fixed galaxy density of $10^{-3}(h/\rm{Mpc)}^{3}$ close to that expected for the DESI ELG sample. In the following, all our results are expressed as a function of the initial (i.e. unrescaled) value of $A_s$, corresponding to $A_c=0.05$.

Throughout the paper, we complement the above functions for the mean numbers of central and satellite galaxies by the following assumptions.
The actual number of central (resp. satellite) galaxies per halo of mass $M$ follows a Bernoulli (resp. Poisson) distribution with mean equal to $\left\langle N_{\mathrm{cent}}(M) \right\rangle$ (resp. $\left\langle N_{sat}(M)\right\rangle$).
Central galaxies are positioned at the center of their halos while satellite galaxy positions sample a Navarro-Frenk-White profile~\cite{NFW}. We assume that satellite velocities are normally distributed around their mean halo velocity, with a dispersion equal to that of the halo dark matter particles, rescaled by an extra free parameter denoted $f_{\sigma_v}$.

\section{Simulation, mock catalogues and clustering statistics}
\label{sec:mock}
 Our HOD fitting method uses the above GHOD model to populate simulated dark matter halos and produce mock galaxy catalogues for which clustering statistics are calculated and compared to data. In this paper, we test our method with simulated data (dubbed as pseudo-data in the following) that are themselves galaxy mock catalogues produced in the same way. For both purposes, we rely on the \absum suite of high-accuracy cosmological N-body simulations~\cite{AbacusSummit} and the corresponding cleaned halo catalogues obtained with the \compaso algorithm~\cite{CompaSO}. This simulation suite was produced for the clustering analyses of DESI. Defined primarily on the base Planck 2018 $\Lambda {\rm CDM}$ best-fit cosmology~\cite{Planck2018} but offering also several variants, the suite contains different resolutions and cubic box sizes. 
 
 Table~\ref{tab:simulation} describes the subset of simulations used in this work. They all use the base resolution, 6912$^3$ particles in a box of 2 ${\rm Gpc}/h$ length, which corresponds to a particle mass of about 2$\times 10^9 M_{\odot}/h$. With this particle mass, halos are well resolved down to $10^{11}M_{\odot}/h$ which provides $\sim$50 particles/halo~\cite{AbacusSummit}. Moreover, the halos selected in this work have a mass larger than $3\times10^{11}M_{\odot}/h$ which corresponds to 150 particles/halo.
Simulations in Table~\ref{tab:simulation} are used to create mock catalogues for both pseudo-data and model predictions, as will be detailed in the next sections. For pseudo-data mocks, the GHOD parameters are fixed at values listed in Table~\ref{tab:pseudo-data} (top row). These values provide a clustering close to that expected for the ELG sample collected during the survey validation phase of DESI.

As clustering statistics, we adopt the projected correlation function, 
which is robust against redshift-space distortions at small scales,
and the two-point correlation function monopole and quadrupole. We first compute the galaxy two-point correlation function, $\xi(r_p,\pi)$, as a function of the galaxy pair separation components along ($\pi$) and perpendicular to the line-of-sight ($r_p$). 
Integration over the line-of-sight provides the projected correlation function:
\begin{equation}
w_p(r_p) = 2 \int_0^{\pi_{max}} \xi(r_p,\pi) d\pi.
\end{equation}
Computing the two-point correlation function $\xi(s,\mu)$, as a function of the galaxy pair separation, $s$, and  the cosine of the angle between the line-of-sight and separation vector, $\mu$, provides the two multipoles we use:
\begin{equation}
\xi_{\ell}(s)= \frac{2\ell+1}{2}\int_{-1}^1 \xi(s,\mu) {\it L}_{\ell}(\mu) d\mu
\end{equation}
with $\ell \in \{0, 2\}$ and where ${\it L}_\ell(\mu)$ denotes the Legendre polynomial of order $\ell$.
We rely on the DESI wrapper (see appendix~\ref{app:app1}) around the \textsc{Corrfunc} package~\cite{Corrfunc} to compute the above two-point correlation functions $\xi(r_p,\pi)$ and $\xi(s,\mu)$
with the natural estimator, which compares galaxy pair counts to the expected pair count for a uniform distribution.
For $w_p(r_p)$, we use 25 logarithmic bins in $r_p$ between 0.03 and 30 Mpc$/h$, setting $\pi_{max} = 40\, {\rm Mpc}/h$. For the multipoles, we use 25 logarithmic bins between 0.8 and 30 Mpc$/h$ in $s$ and 100 linear bins in $\mu$. In the galaxy pair count computation, the galaxy redshift to distance conversion uses the simulation cosmology as the fiducial cosmology and the $z$ axis is chosen as a line-of-sight for the application of redshift space distortions.

\begin{table}[tbp]
\centering
\begin{tabular}{l|c|c|c|c|}
  name  &  cosmology & box size &  resolution & realisations \\ \hline
  baseline & Planck 2018 $\Lambda_{\rm CDM}$ & 1 ${\rm Gpc}/h$ & 3456$^3$ & 1 \\
  cosmic variance & Planck 2018 $\Lambda_{\rm CDM}$ & 2 ${\rm Gpc}/h$ & 6912$^3$ & 25 \\
\end{tabular}  
\caption{\label{tab:simulation} Cosmology, box size and mass resolution of the \absum simulations used in this work. The mass resolution is given as the number of particles in the box.
}
\end{table} 

\begin{table}[tbp]
\centering
\begin{tabular}{l|c|c|c|c|c|c|c|}
parameter   &  $\log_{10}(M_c)$ &  $\alpha$ & $A_s$ & $\log_{10}(M_0)$ & $\log_{10}(M_1)$ & $\sigma_{\textsc{m}}$ & $f_{\sigma_v}$ \\ \hline  
input  & 11.63 & 0.6 & 0.11 & 11.63 & 11.93 & 0.12 & 1.\\
priors   & 11.4-11.8 & 0.5-0.7 & 0.05-0.2 & 11.4-11.8 & 11.93 & 0.01-0.3 &  0.75-1.25 
 \end{tabular}
 
\caption{\label{tab:pseudo-data} {\it Top row.} GHOD parameter input values used for pseudo-data catalogues. {\it Bottom row.}  GHOD parameter flat prior ranges used in all fits performed in this paper. 
In both rows, the indicated values of $A_s$ are initial values and the initial value for $A_c$ is 0.05. We recall that, at each point in the parameter space, these individual values are rescaled by the same factor to provide clustering modelling with a fixed density of $10^{-3} (h/{\rm Mpc})^3$ (see text).
}
\end{table} 

\section{Inference methodology}
\label{sec:methodology}
The methodology we present hereafter aims at performing accurate fitting of HOD model parameters while minimising CPU time consumption. 
To this purpose, inspired by Efficient Global Optimization algorithms \cite{Jones98}, we developed a two-step procedure using Gaussian Processes (GP) to create a surrogate model of the likelihood posterior. In a first step, we sample 
the likelihood posterior to provide initial training to the GP. This initial GP model is further improved by successive iterations, each iteration adding one point until the predicted map becomes stable enough so that marginalised parameter values and posterior contours can be reliably derived.
After a brief introduction to Gaussian processes, we describe the different steps of the fitting procedure and give its performance.

\subsection{Gaussian Processes}
The interested reader is referred to~\cite{Pellejero-Ibanez20, GPry_2022} for a clear and detailed account of Gaussian processes as an efficient emulation technique. 
In the following, the description is kept to a minimum.
Gaussian processes provide a way to predict the likelihood for any set of parameter values, from the likelihood values computed for an initial, restricted set of parameter values (called training data in the following). To do so, the probability distribution of the prediction, knowing the computed values, is derived assuming that the computed and predicted likelihood function values are jointly distributed Gaussian variables. 
As often assumed in Gaussian processes, the mean value of the joint Gaussian distribution is assumed to be 0. 
The covariance matrix of the joined distribution is specified by a kernel function, to be chosen by the user. We performed tests with a squared exponential kernel (also known as Radial Basis Function, RBF) and a Mat\'ern kernel of index $5/2$ which is equivalent to the product of an exponential and a polynomial of order 5. We adopt the latter as our baseline for this work.

The free parameters of the kernel are one length scale for each input parameter. Initially set to one and allowed to vary in the range between $10^{-3}$ and 10, their values are found by optimisation during the learning process based on the training data. Likelihood measurement uncertainties are included in the procedure, as described in the next section.

\subsection{$\chi^2$ definition}
\label{sec:chi2}

At each point of the HOD parameter space, 20 model realisations are drawn. For each realisation, the model clustering is compared to that of the pseudo-data with the following $\chi^2$ definition: 
\begin{equation}
\chi^2 = (\vect{\xi}_{data}-\vect{\xi}_{model})^\top
\big[\vect{C}_{data}/(1-D_{data})+\vect{C}_{model}/(1-D_{model})\big]^{-1}
(\vect{\xi}_{data}-\vect{\xi}_{model})
\end{equation}
where $\vect{\xi}$ denotes a vector of clustering measurements, $\vect{C}$ the corresponding covariance matrix and $D$ the Hartlap correction factor~\cite{Hartlap}. 
These 20 measured $\chi^2$ values are then averaged
and the dispersion of the $\chi^2$ values divided by $\sqrt{20}$ is used as an estimate of the uncertainty on the mean $\chi^2$. For HOD input values in Table~\ref{tab:pseudo-data} (top row), this uncertainty is of order 2.3 for a mean $\chi^2$ around 63 (for $75-6=69$ degrees of freedom). 
As the dynamical range of $\chi^2$ variations is large over the HOD parameter space, which can make it hard to model the likelihood posterior, we use the natural logarithm of the mean $\chi^2$ values and the corresponding errors as inputs to the GP.

The computation of the covariance matrix for the pseudo-data depends on the test to be performed and is described in the next section. 
To build the model covariance 
we assume that correlations have small variations over the HOD parameter space
and compute a fixed correlation matrix from 1000 realisations of the HOD model in table~\ref{tab:pseudo-data} (top row), drawn from the simulation box used for the model. At each point of the parameter space, the model covariance matrix is then obtained by normalising the previous correlation matrix using the variances of the clustering measurements over the 20 realisations drawn to compute the $\chi^2$ at that point.
This is the baseline for the computation of the model covariance matrix, which we changed slightly for specific tests, as detailed in Section~\ref{sec:tests}.

\subsection{GP training sample}
\label{sec:training}

To define the GP training sample, the HOD parameter space must be sampled efficiently. Two sampling methods were tested. The first one, the Latin Hypercube Sampling (LHS) 
partitions the parameter space into bins of equal probability so as to provide a more even distribution of sample points than would be possible with pure random sampling. However, we noticed that LHS can be too sparse for HOD fitting, resulting in biased contours or even missed best fits. The second method, which we adopt as our default in this paper, is the Hammersley sampling~\cite{Hammersley} which generates a more uniformly distributed sampling pattern at low computational cost. This algorithm is reliable and efficient for low-dimensional problems only (less than 10 parameters) which is the case of this work. Note that the Hammersley sampling assumes equidistant points in one dimension of the parameter space. 

To define the training sample
we draw $N=600$ points in the HOD parameter space defined by Hammersley sampling and, in each point, compute the previously defined $\chi^2$ and its error. The GP is then provided with the natural logarithm of the $N$ computed $\chi^2$ values, together with the corresponding errors.
Throughout the paper, the parameter values are drawn uniformly in ranges summarised in Table~\ref{tab:pseudo-data} (bottom row). Unless otherwise stated, we choose $A_s$ as the parameter space dimension with equidistant points.
Since the prior ranges do not change throughout the paper, the Hammersley sampling for a given number of points and a given choice for the dimension with equidistant points is uniquely defined. We study the impact of changing these conditions in section~\ref{sec:tests}.

\subsection{Iterations and fit stability criterion} 
\label{sec:convergence}

\begin{figure}[tbp]
\centering
\includegraphics[width=.9\textwidth]{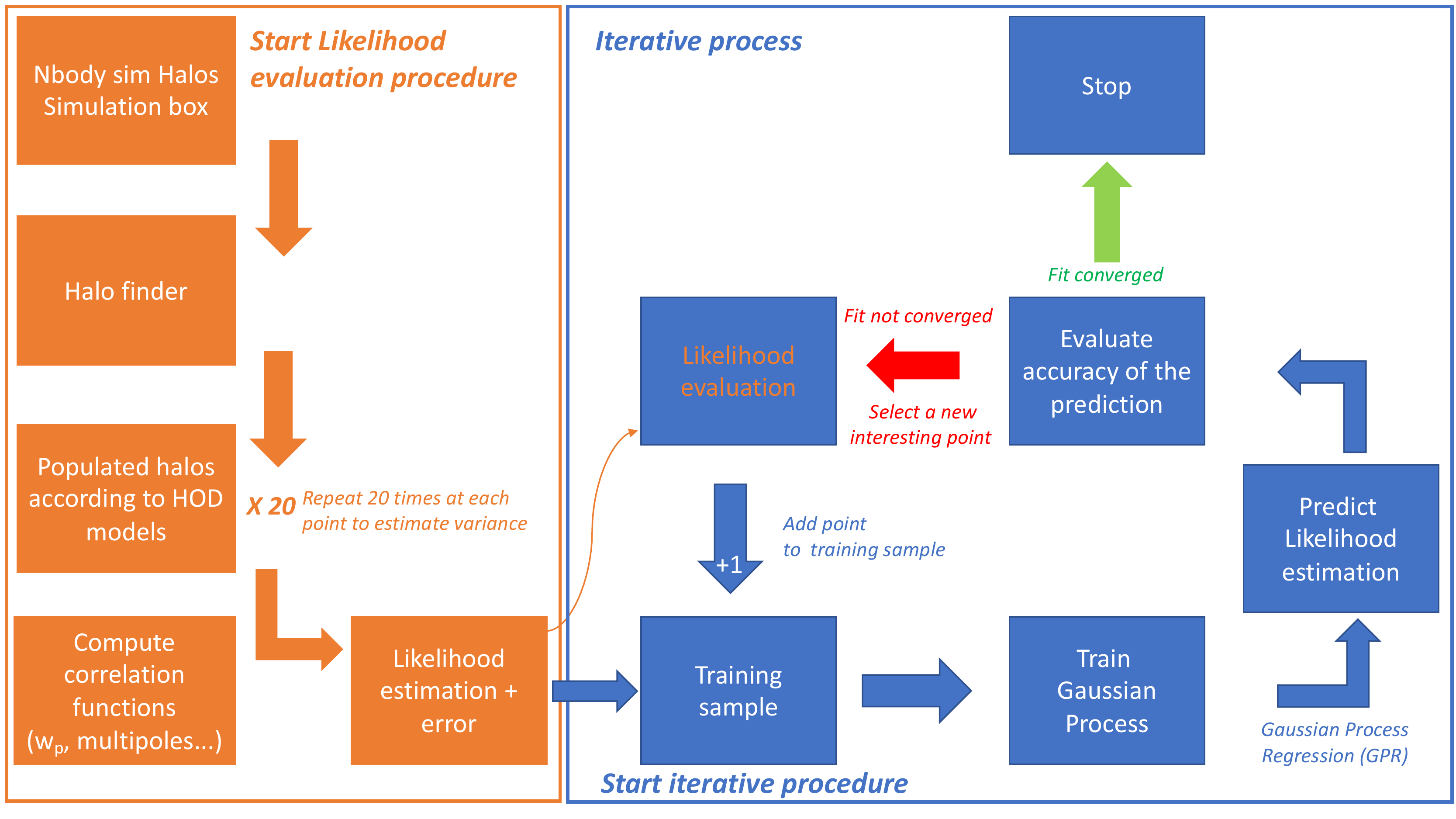}
\caption{\label{fig:sketch} Sketch of the fitting procedure. {\it Left:} mock catalogue creation, clustering measurements and likelihood computation. {\it Right:} fitting iterative procedure based on likelihood predictions driven by Gaussian processes.}
\end{figure} 

The surrogate model of the likelihood posterior provided by the GP from the initial training sample is iteratively improved by adding one point to the training sample at each iteration (see Figure~\ref{fig:sketch} for a schematic view). Choosing the next point to add, $x_{next}$, that is choosing the GP acquisition function,   can be done in several ways. 
The most popular one uses the Expected Improvement (EI) information acquisition function~\cite{Jones98,EI}. The latter defines how much the likelihood value at a given point is expected to improve over the current maximum and the point that gives the greatest expected improvement is taken as $x_{next}$. Applied to our case, this method proved to be efficient at finding the maximum of the likelihood function but did not provide accurate error contours. This illustrates the difficulties to define an acquisition function that reaches a good compromise between exploring the full parameter space and focusing on high probability areas, as discussed in~\cite{GPry_2022}.

In order to have both an accurate determination of the likelihood maximum and reliable error countours, we use the following method to determine $x_{next}$, based on previous works in cosmology \citep{Pellejero-Ibanez20, Neveux21}. At each iteration, the GP prediction is sampled by a Monte Carlo Markov chain (MCMC) algorithm and $x_{next}$ is randomly selected in the MCMC chains. Its $\chi^2$ value and error are computed and the point is added to the training sample to reiterate the procedure. This method has the advantage that the points inside the 3 $\sigma$ contour around the maximum likelihood are more likely to be selected as points to be added to the training sample. 

\begin{figure}[tbp]
\centering
\hspace{-0.3cm}
\includegraphics[width=1\textwidth]{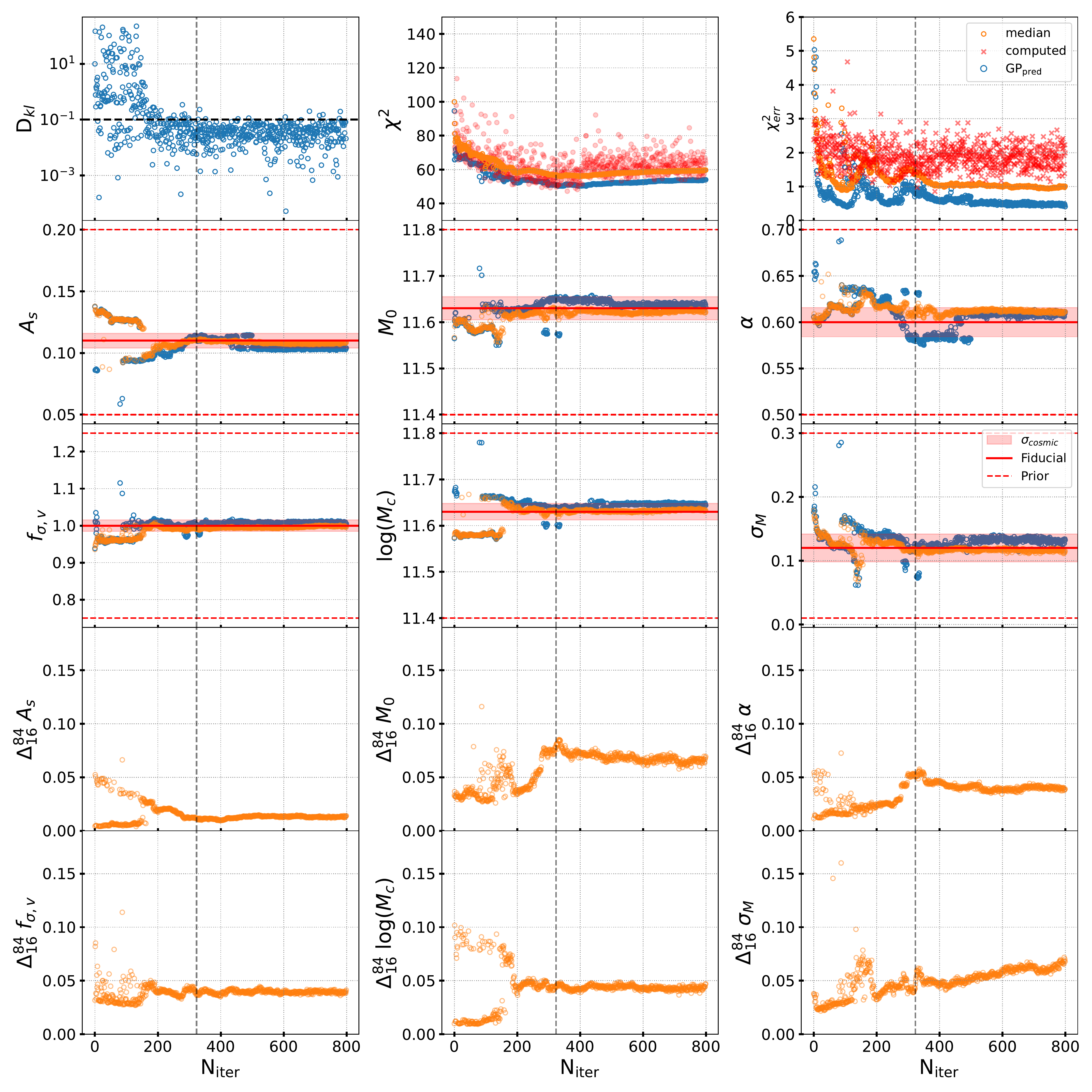}
\caption{\label{fig:convergence} Evolution of several fit result indicators with iteration number, for one fit from the accuracy test with cosmic variance included (see section~\ref{sec:tests}). From left to right and top to bottom, indicators are the Kullback-Leibler divergence between MCMC chains sampling the GP prediction, the value of $\chi^2_{min}$ (blue) and that of $\chi_{med}^2$ (orange) from the chains, their errors, the values of the six HOD parameters at  $\chi^2_{min}$ (blue) and their marginalised median values (orange), the $[16-84]$ percentile range of the six HOD parameters. Red dots in the middle (resp. right) top panel are $\chi^2$ values (resp. errors) computed at the selected point added to the GP training sample at the next iteration. Red solid (resp. dashed) lines are fiducial (resp. prior boundary) values. The band indicates the dispersion from all fits with cosmic variance included. The vertical dashed line indicates the iteration at which the KL criterion is met.
}
\end{figure} 

To stop the iterative procedure, we require stability of both the MCMC chains and fit results, as explained in the following.
To characterise the chain stability, we rely on     the Kullback-Leibler (KL) divergence \cite{KL} between the MCMC chains, computed at each iteration. We consider our KL criterion as fulfilled when the KL divergence is below 0.1 in a set of 20 consecutive iterations before the current one. This threshold was determined empirically from our fits, but is rather common in iterative GP emulators~\cite{Pellejero-Ibanez20, Neveux21}.
To illustrate the stability of the chains and fit results, Figure~\ref{fig:convergence} shows an example of the evolution of several indicators of the fit results as a function of the iteration number.

The fit results on HOD parameters at each iteration are characterized both by values corresponding to the minimum $\chi^2$ value of the MCMC chains run on the GP prediction (hereafter called $\chi^2_{min}$), and by marginalised values defined as the median values of the posterior distributions run from the same MCMC chains. Errors on these marginalised values are defined by the $16\%$ and $84\%$ percentiles of the parameter posterior distributions.
Besides the KL divergence, the fit results reported in Figure~\ref{fig:convergence} are thus the values of the six HOD parameters at  $\chi^2_{min}$ and their marginalised values, as well as the value of $\chi^2_{min}$ and the $\chi^2$ value for the GP prediction at the marginalised HOD parameter values (hereafter dubbed as $\chi^2_{med}$), together with their corresponding errors. We also show the evolution of the size of the $[16-84]$ percentile range of the posterior distribution for the six HOD parameters. 
Finally, in the sub-plots related to $\chi^2$ values and errors, we added the computed $\chi^2$ value and its error for the point added at each iteration.

\begin{figure}[htbp]
\centering
\begin{tabular}{c}
\includegraphics[width=0.9\textwidth]{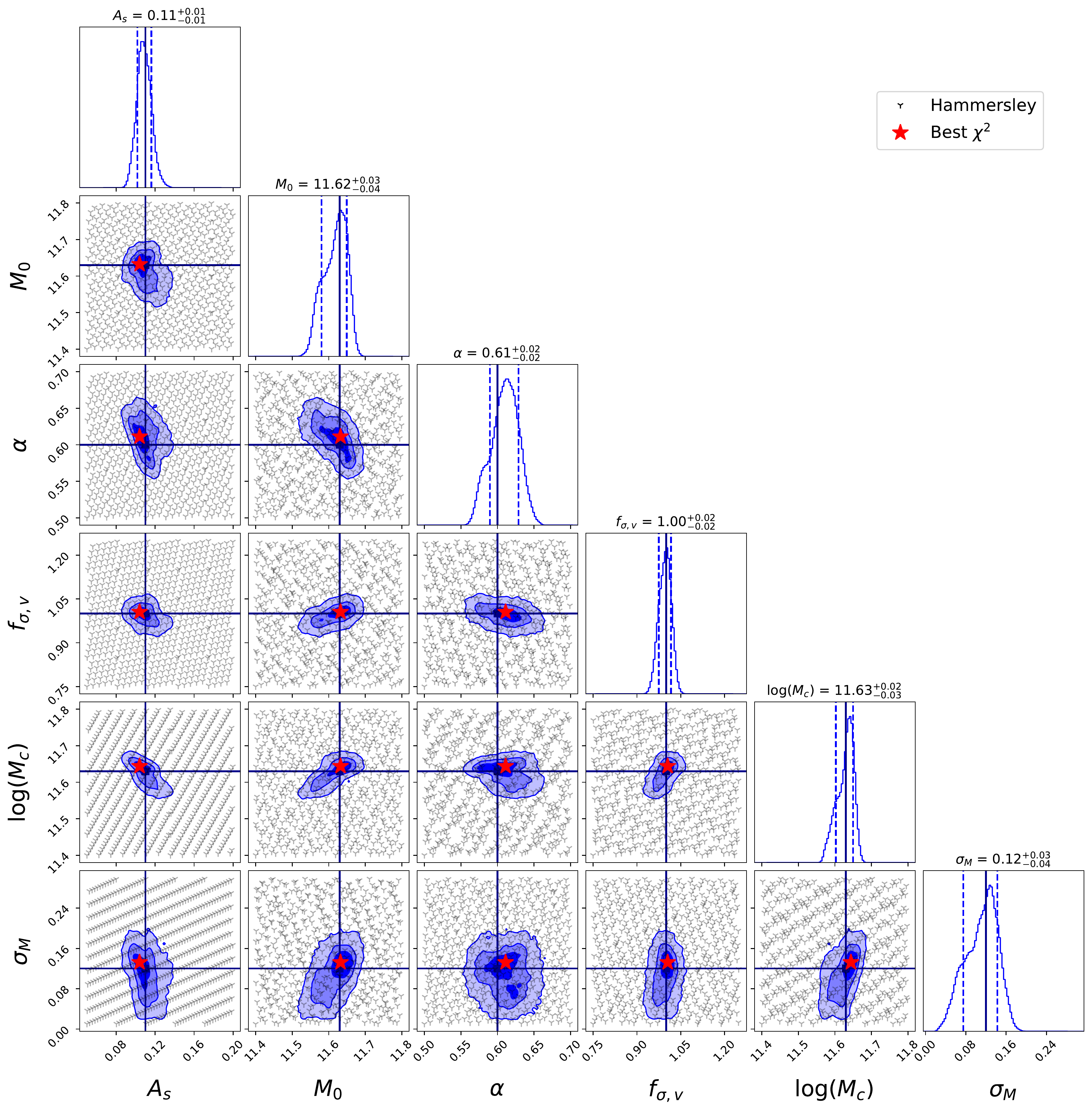} \\
\hspace{-0.3cm}
\includegraphics[width=1\textwidth]{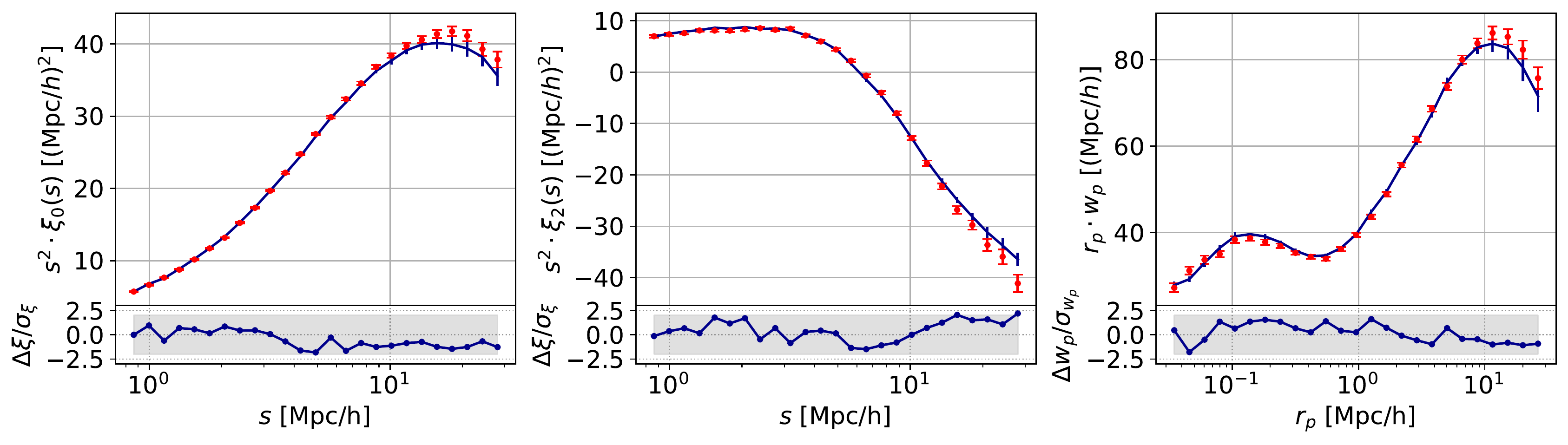}
\end{tabular}
\caption{\label{fig:ite800} {\it Top.} Contours and marginalised 1D-posteriors at iteration 800 for the fit in figure~\ref{fig:convergence}. Grey points are the Hammersley training sample. The red star corresponds to the minimal $\chi^2$ of the GP prediction. The contours are obtained from MCMC chains run on the GP predictions, after burning phase. Solid lines indicate the parameter input values. 
{\it Bottom.} Clustering measurements predicted by the HOD model 
from the fit 
in the top plot. The shaded band corresponds to $\pm2\sigma$ residuals, with errors on pseudo-data (red bars) and model (blue bars) added in quadrature.
}
\end{figure}

As can be seen from the figure, the KL divergence drops and remains below the threshold of $0.1$ after iteration 300. The values of $\chi^2_{min}$ and $\chi^2_{med}$ reach a plateau after that iteration but 
the learning phase of the GP continues, as shown by the excursions in the $\chi^2_{min}$ error. The explored range in the parameter values, indicated by the excursions of their values at $\chi^2_{min}$
are limited and induce small variations of the marginalised HOD parameter values and their percentile ranges, which are all well stabilized at iteration 800, although the percentile range of $\sigma_M$ may still be evolving slightly.
For completeness, Figure~\ref{fig:ite800} presents the contour plot of the fit at iteration 800 (see also Appendix~\ref{app:conv}) and compares the pseudo-data clustering to that from the HOD model defined by the marginalised parameter values at that iteration. The modelling of the pseudo-data clustering is good, well within errors due to model stochasticity and cosmic variance, the latter being the dominant effect at large scales. The $\chi^2_{med}$ value 
at the final iteration is $56.8 \pm 2.2$ for a number of degrees of freedom of $75-6=69$ (p-value of 85$\%$). 

Although the fit in the above example reaches stable and reliable results after 200 iterations are added once the KL criterion is met, nothing guarantees that this is generally the case~\cite{GPry_2022}. The stability of the results from the iterative method  
will thus be investigated in a systematic way in section~\ref{sec:tests}, by running the procedure on a large set of simulated mocks, taking into account cosmic variance.
All fits will be run up to a maximal number of iterations $N_{max}=800$
and the fit precision and accuracy of the method will be explored with that stopping criterion, which we discuss further in section~\ref{sec:tests}.
The fit results will be defined by the marginalised HOD parameter values at  iteration $N_{max}$, with statistical uncertainties on these given by the $16\%$ and $84\%$ percentiles of the parameter posteriors at that same iteration.

\subsection{Performance}
Implementation details of the above procedure are described in appendix~\ref{app:app1}. Its performance are as follows. 

Our computer system uses two 
AMD EPYC 7513 32-core processors, which are multi-threaded by 2 (128 threads/node in total), clocked at 2.6~GHz and equipped with 256~GB DDR4 RAM. The tests described in this paper were run with 24 threads on a single node. The CPU time consumption per point of the HOD parameter space breaks down as follows: $\sim 25$ sec to create 20 realisations of the HOD model from a cubic simulation box of 1~$\mathrm{Gpc}/h$ length, $\sim 12$ sec to compute the correlations and the $\chi^2$ value based on these 20 realisations, $\sim 20$ sec to run the MCMC chains on the GP prediction. The CPU time consumption to derive the prediction of the GP depends on the number of points in the training sample. 
Altogether, for a 6-parameter fit based on 600 training points and 800 added points, the total CPU time consumption per iteration increases from $\sim 50$ seconds with the initial training to $\sim 3$ minutes at the final iteration.

\section{Tests of the method}
\label{sec:tests}
We first test to which extent the procedure can be considered as reproducible and then include cosmic variance to test the method accuracy. We then study how the results evolve when the ingredients of the method are changed.

\subsection{Reproducibility}
\label{sec:repro}
To test the reproducibility of the method, we use the 1 Gpc$/h$ length cubic box in the base cosmology (see Table~\ref{tab:simulation}) to create one pseudo-data mock with the HOD parameter input values in Table~\ref{tab:pseudo-data} (top row). The data covariance matrix is computed from 1,000 realisations of the same HOD model on the same box. 
This covariance matrix includes stochastic noise in the process of populating halos with galaxies and the statistical noise induced by the density of the resulting mock catalogues, the two irreducible sources of noise of the procedure.
Data and model variances were compared (for the input HOD model) and found to be comparable in all separation bins of the clustering measurements. 

\begin{figure}[tbp]
\centering
\includegraphics[width=\textwidth]{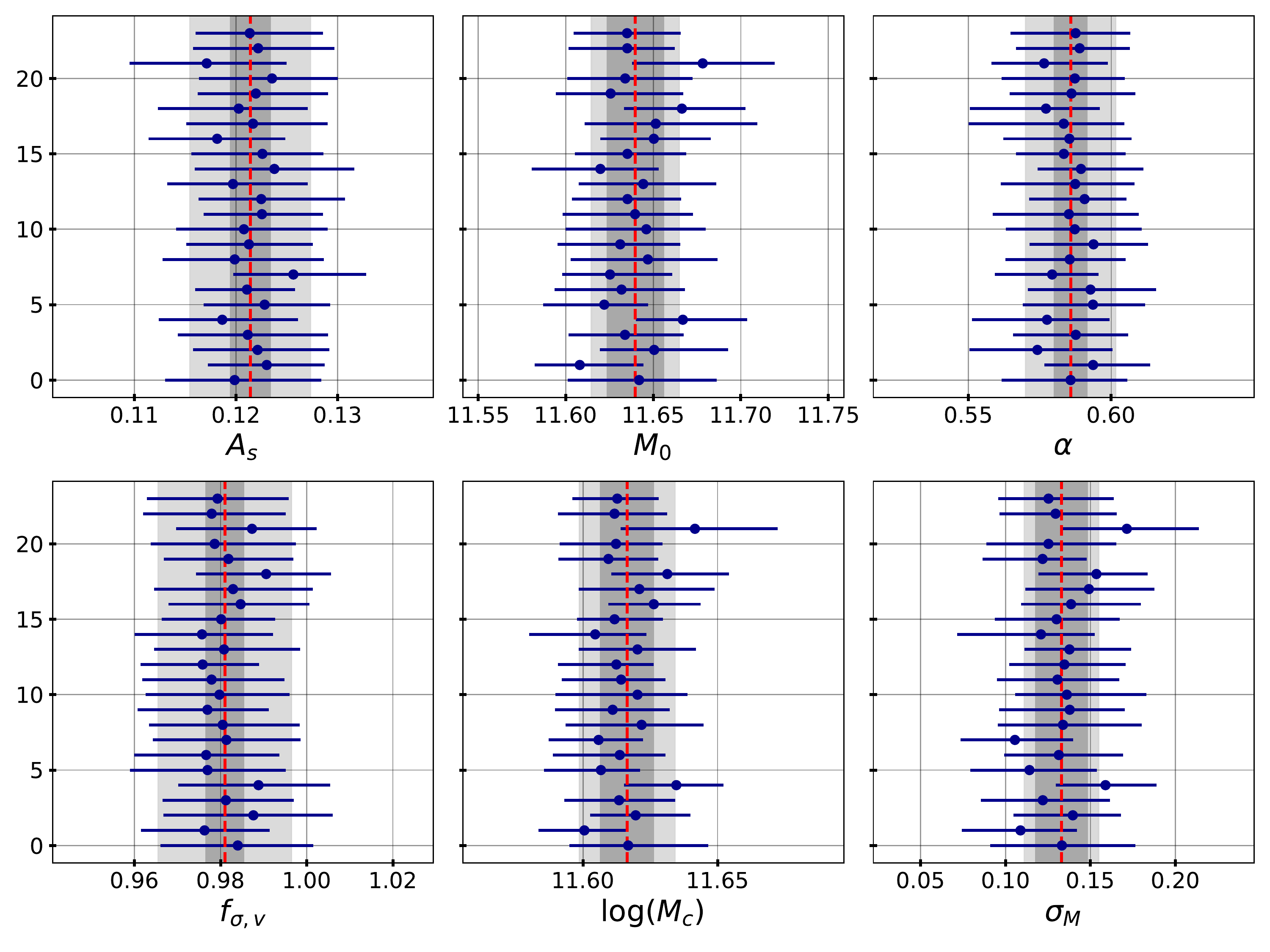}
\caption{\label{fig:repro} Reproducibility test from 25 fits to the same pseudo-data under the same initial conditions.  Blue dots are marginalised HOD parameter values with errors defined by the $16\%$ and $84\%$ percentiles of the fit posteriors. The dark grey band spans the $\pm 1\sigma$ interval around the average of the marginalised values given by the vertical red dashed line. The light grey band includes also cosmic variance (from the accuracy test of section~\ref{sec:accuracy}).
}
\end{figure}

We perform 24 independent fits to the pseudo-data mock, all with the same initial sampling of the parameter space (see section~\ref{sec:training}). Results are presented in Figure~\ref{fig:repro}. The results of the 24 fits agree with each other within $\pm0.002$ for $A_s$, $\pm0.005$ for $\alpha$, $\pm0.004$ for $f_{\sigma_v}$, $\pm 0.013$ for $\sigma_{\textsc{m}}$ and within $\pm0.016$ and $\pm0.010$ for $M_0$ and $\log_{10}(M_c)$ respectively. 
These numbers quantify the reproducibility limits at the 68$\%$ confidence level of our procedure, due to its stochastic nature. Also included in the plot are the expected dispersions when cosmic variance is also taken into account (see next section), showing that, except for $M_0$ and $\sigma_{\textsc{m}}$, 
the intrinsic dispersion due to stochasticity is a sub-dominant component. 

Spurious instabilities in the GP predicted likelihood posterior were observed for these fits but were overcome with a simple post-processing treatment, removing points with large GP predicted $\chi^2$ errors from the MCMC chains, as
explained in appendix~\ref{app:repro}. 

\subsection{Accuracy with cosmic variance}
\label{sec:accuracy}
To test the method accuracy in more realistic conditions, cosmic variance must be included. We thus cut each of the 25 large boxes of 2 Gpc$/h$ length in the base cosmology (see Table~\ref{tab:simulation}) into 1~Gpc$/h$ length cubes, which allows us to create 200 independent mocks corresponding to different realisations of the same cosmology. 

We take 25 out of these 200 mocks as pseudo-data (one per large box) and perform four series of fits, each series using a different 1~Gpc$/h$ sub-cube for the model. These sub-cubes differ from those used to create the 25 pseudo-data mocks and three of them belong to the same large box. The training sample 
is recomputed for each of the 4 sub-cubes.

The data covariance matrix is built from the entire set of 200 mocks. This matrix includes stochastic noise in
the process of populating halos with galaxies, statistical noise induced by mock density, and cosmic variance. The model covariance matrix is built as described in Section~\ref{sec:chi2}, but the variances used to compute the normalisation factor applied at each HOD point to the fixed correlation matrix are modified to account also for cosmic variance in the choice of a given sub-cube for the model. To the variances used in Section~\ref{sec:chi2} to define the normalisation factor, we thus add the clustering measurement variances over the 8 sub-cubes of the box used for the model, computed for the input HOD, with one realisation per sub-cube.
Data and model variances were compared and their difference  was found to be within $\pm 0.6$ times the pseudo-data variances, with no marked scale dependence.

\begin{figure}[tbp]
\centering
\includegraphics[width=\textwidth]{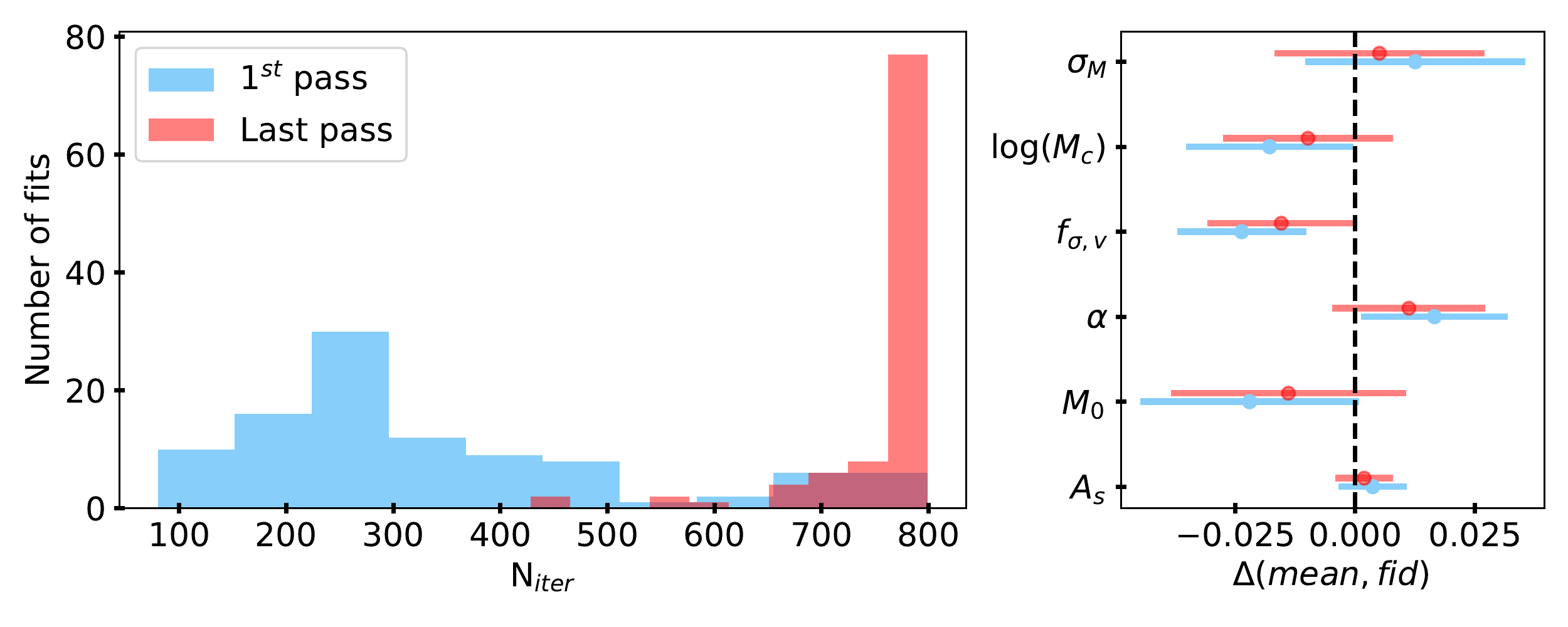}
\caption{\label{fig:diff_1st_last} \textit{Left.} Distribution of first (in light-blue) and last (in red) iteration fulfilling the Kullback-Leibler stability criterion for a set of 100 fits with cosmic variance included. \textit{Right.} Difference between the fiducial values and the mean results of the 100 fits, for each of the 6 HOD parameters. Results from fits stopped at the first (resp. last) iteration fulfilling the Kullback-Leibler criterion are reported in light-blue (resp. red). The error bars correspond to the $\pm$ one standard deviation of the 100 marginalised fit results. 
All fits had initial training of the Gaussian Processes based on Hammersley sampling of the HOD parameter space with 600 points.
}
\end{figure}



\begin{figure}[tbp]
\centering
\includegraphics[width=\textwidth] {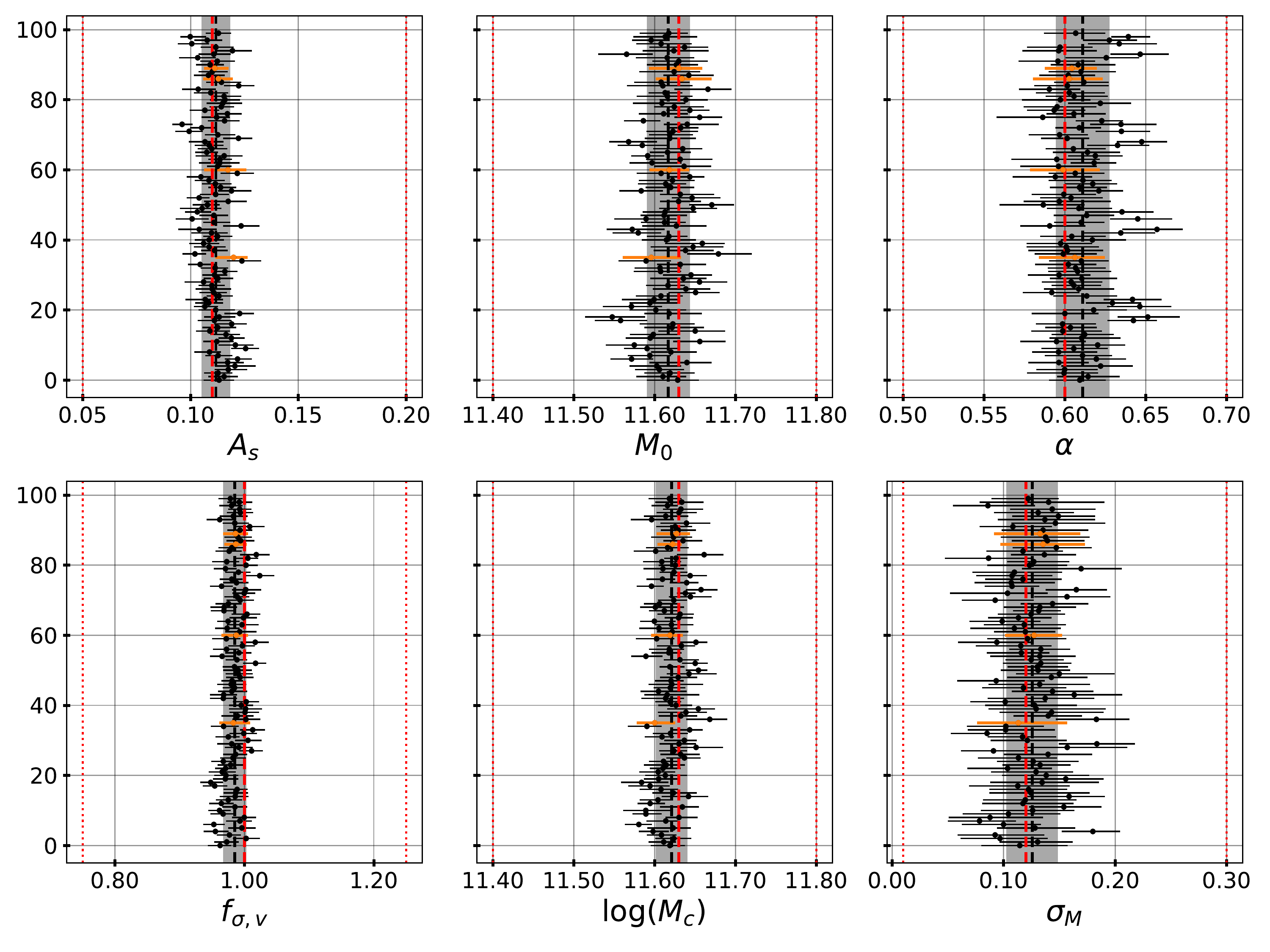}
\caption{\label{fig:accu} Accuracy test from 100 fits with cosmic variance included. Dots are marginalised HOD parameter values with errors defined by the $16\%$ and $84\%$ percentiles of the fit posteriors. Orange dots stand for the four fits which did not converge. The grey band spans the $\pm 1\sigma$ interval around the average of the marginalised values given by the vertical black line. The red dashed line is the input HOD parameter values.
Four series of fits to the same 25 pseudo-data mocks were run with a model drawn from a different sub-cube of the same large box for the first three series and from a sub-cube of a different large box in the fourth one. All fits were run up to 800 iterations after initial training of the Gaussian Processes based on Hammersley sampling of the HOD parameter space with 600 points. Red dotted lines indicate the fit priors. 
}
\end{figure}

The distribution of the first iteration meeting the KL criterion as defined in section \ref{sec:convergence}, is presented in Figure~\ref{fig:diff_1st_last} (left-hand plot, blue histogram). 86\% of the fits pass the KL criterion before iteration 600 and only 4$\%$ did not reach stability at the last iteration, $N_{max}=800$.
Marginalised fit results are compared to the input HOD parameter values in Figure~\ref{fig:accu}. The four fits which did not pass the KL criterion do not stand as clear outliers in any of the parameters, showing that the lack of stability does not necessarily imply a large offset in the measured parameters, nor larger error bars. 
Figure~\ref{fig:diff_1st_last} also shows the distribution of the last iteration meeting the KL criterion (left-hand plot, red histogram), which is strongly peaked towards $N_{max}$. The right-hand plot in this figure compares the bias observed on each parameter when the fits are stopped at the first or last iteration meeting the KL criterion, instead of allowing all fits to go up to $N_{max}$. Stopping the fits at the first iteration reaching the KL stability criterion clearly leads to larger biases than stopping at the last one 
or going up to $N_{max}$ (biases in these two cases are very similar), which justifies our choice of the latter option as a stopping criterion. 

With this criterion, all HOD parameters are reconstructed with a mean bias either well within, or for $f_{{\sigma}_v}$ at the level of, one standard deviation of the parameter distribution. More precisely, we find a mean bias of $0.29\sigma$ for $A_s$,  $0.52\sigma$ for $M_0$, $0.69\sigma$ for $\alpha$, $0.97\sigma$ for $f_{{\sigma}_v}$, $0.52\sigma$ for $\log(M_c)$ and $0.26\sigma$ for $\sigma_M$.
In the above, $\sigma$ is the standard deviation of the marginalised fit result distribution. This is also the expected statistical error for one fit, with the errors accounted for in the covariance matrix used in the fits, namely stochasticity, cosmic variance and galaxy sample size for a density close to that expected for DESI data but for a volume three times larger than that of the early DESI ELG data. 
We thus expect the HOD parameter values to be derived from these data with our procedure to have an accuracy much better than 1~$\sigma$ of the data statistical uncertainty for most parameters, the worst case being the $f_{{\sigma}_v}$ parameter for which the accuracy is expected to be about 0.6~$\sigma$.

The above reasoning assumes statistical errors from the fits to be normally distributed. As a cross-check, we compared the mean of the parameter errors from the fits to the standard deviation of the marginalised fit result distribution used in the above bias estimates. 
We found the mean error to be higher than the standard deviation by 60$\%$ for $\sigma_M$ and 20-30$\%$ for all other parameters. These departures are likely to be due to non Gaussian posteriors, as observed in most parameters (see Figure~\ref{fig:ite800}), and make our expected bias estimates conservative. 

However, despite the slight biases observed in the HOD parameters, the procedure provides a very good modelling of the clustering statistics, as already shown in Figure~\ref{fig:ite800} on one example. Altogether, in the four series of fits to the 25 pseudo-data mocks used in this section, the mean value of the computed reduced $\chi^2$ for the best fit model (defined by the marginalised HOD parameter values) is $\sim$0.8 with a dispersion of $\sim$0.15.

\subsection{Dependence on initial conditions and kernel}

In this section, we test how the results of our procedure are affected by different initial conditions. We present results from different numbers of points in the initial training sample, from different GP kernels and different initial sampling algorithms. The priors being unchanged, testing different numbers of points in the training phase amounts to testing different densities when sampling the HOD parameter space.

\subsubsection{Initial training sample}

\begin{figure}[tbp]
\centering
\includegraphics[width=\textwidth]
{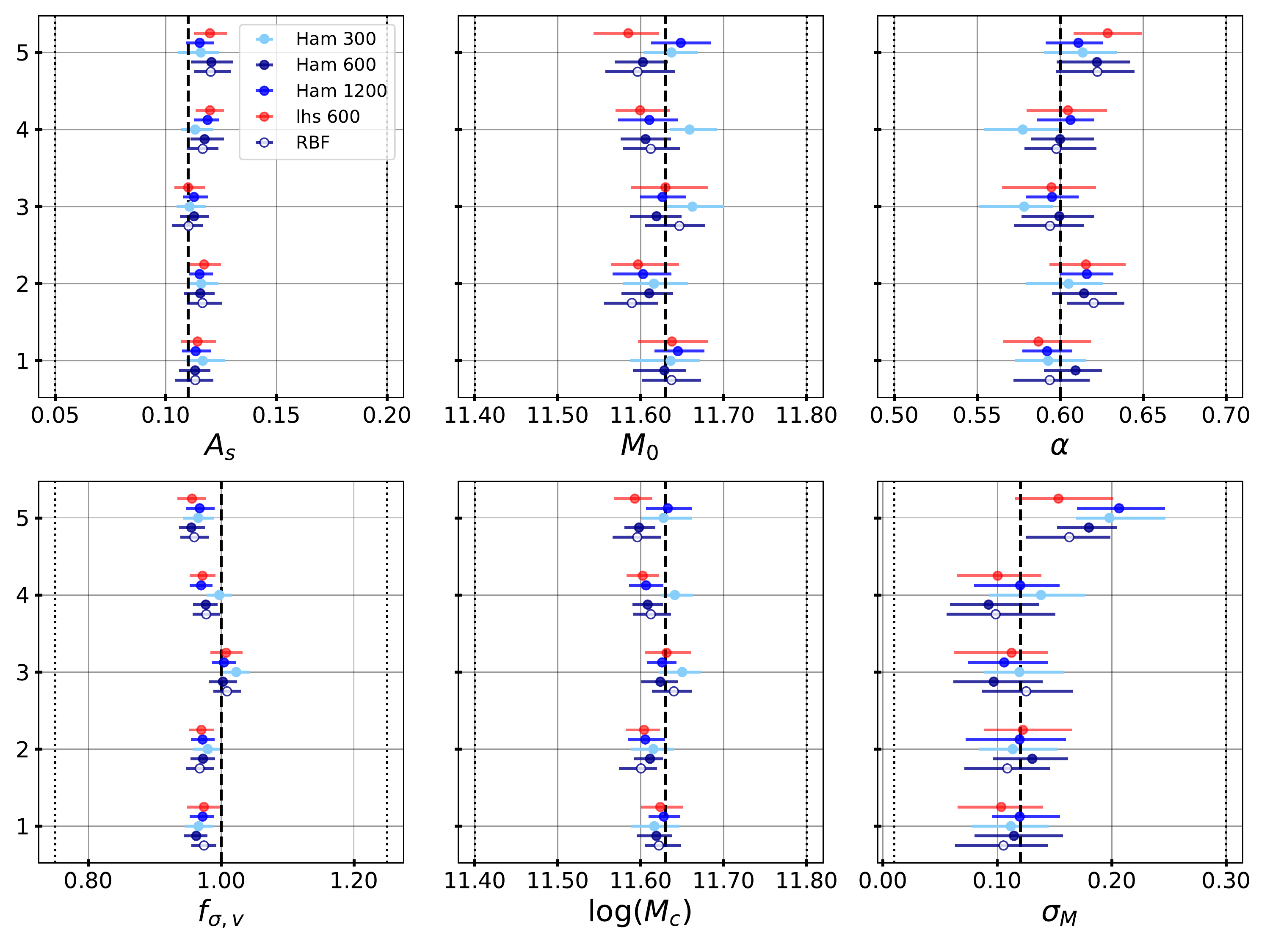}
\caption{\label{fig:nb_point_diff} 
Tests from fits with different training samples and different kernels for the Gaussian Processes. Five series of fits to the same 5 pseudo-data mocks with the same box for the model were run with different sizes of the Hammersley training sample (from 300 to 1200), or replacing Hammerlsey sampling by LHS for the baseline sample size (600) or using a Radial Basis Function kernel instead of the baseline Mat\'ern kernel of index 5/2 for the baseline sample size (600).
All fits were run up to 800 iterations.
Dots are marginalised HOD parameter values with errors defined by the $16\%$ and $84\%$ percentiles of the fit posteriors. 
The black dashed line is the input HOD parameter values and the dotted ones indicate the fit priors.  
}
\end{figure} 

Our baseline option for the initial training sample is 600 points distributed according to Hammersley sampling, with $A_s$ as the parameter with equidistant points. Using the first 5 pseudo-data mocks submitted to fits with cosmic variance (see section~\ref{sec:accuracy}), we run fits with Hammersley training samples of 300 and 1200 points, and with LHS training samples of 600 points. All other fits conditions remain unchanged, notably running all fits up to 800 iterations, with the same box for the model.

Results are reported in figure~\ref{fig:nb_point_diff}. The dispersion of the results between different conditions of fits for a given pseudo-data mock is generally lower than the dispersion between mocks for the same fitting conditions, which is dominated by cosmic variance. We note however that sampling with only 300 points appears to gives results less consistent with fits in other conditions for $M_0$, $\alpha$ and $\log(M_c)$. LHS sampling gives slightly larger errors than Hammersley sampling with the same number of points and there is no obvious gain in accuracy with 1200 points compared to 600 points in Hammerlsey sampling. 

This means that there is no need to increase the density of the initial sampling infinitely, at some point what matters most is to increase the density in the region of interest, close to the likelihood maximum, which is the aim of the iterative procedure that follows initial sampling.

\subsubsection{Choice of GP kernel}

In a second test, the 5 pseudo-data mocks were submitted to fits  with Hammersley sampling of 600 points and $A_s$ as the parameter with equidistant points but with an RBF kernel instead of the baseline one, a Mat\'ern kernel of index 5/2. Again, all other fits conditions remained unchanged.
Results are reported in figure~\ref{fig:nb_point_diff}. Changing the kernel does not change the results significantly. We note that the RBF kernel leads to slightly larger uncertainties on the parameters, the average increase ranging from 4$\%$ for $A_s$ to 23$\%$ for $\log(M_c)$. 

\subsubsection{Choice of parameter with equidistant points}

\begin{figure}[tbp]
\centering
\includegraphics[width=\textwidth]{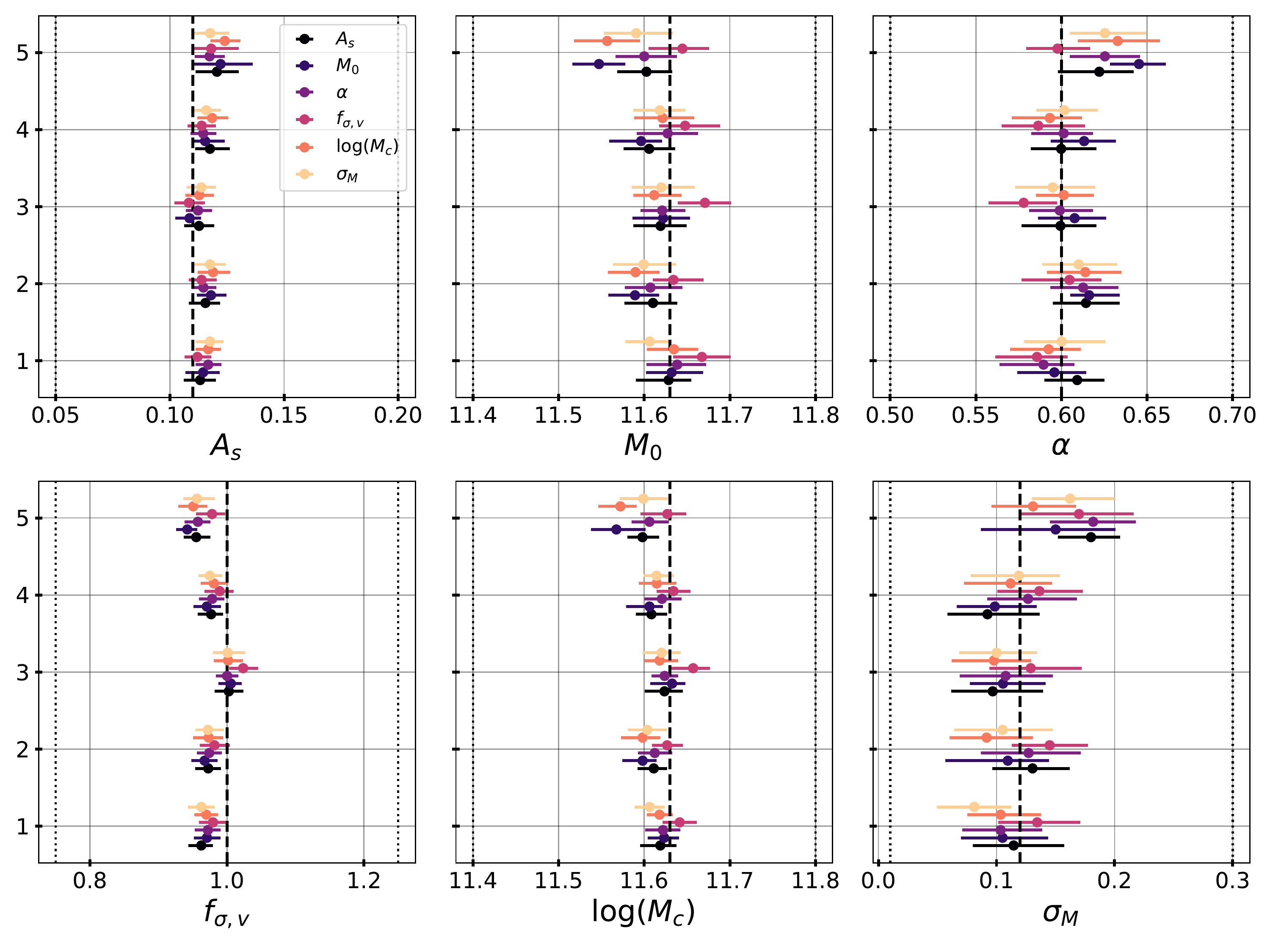}
\caption{\label{fig:1st_par_diff} 
Tests from fits with different training samples. Four series of fits to the same 5 pseudo-data mocks with the same box for the model were run with 600 points from Hammersley sampling, varying the parameter with equidistant sampling. 
Dots are marginalised HOD parameter values with errors defined by the $16\%$ and $84\%$ percentiles of the fit posteriors. Different colors indicate different choices for the parameter with equidistant sampling, $A_s$ being the default for this paper.
The black dashed line is the input HOD parameter values and the dotted ones indicate the fit priors.
}
\end{figure} 

In a third test, the same 5 pseudo-data mocks were submitted to fits  with Hammersley sampling of 600 points, varying the parameter with equidistant points, all other fitting conditions remaining unchanged.
Results are reported in figure~\ref{fig:1st_par_diff}. 
Except for pseudo-data mock 5, the dispersion of the results between the different choices for a given pseudo-data mock is small, but we note that choosing $f_{\sigma_v}$ as the parameter with equidistant points 
can give results differing by $1 \sigma$ from those with other choices, notably for $M_0$.
 
Finally, as already observed in section~\ref{sec:accuracy}, there is a slight systematic offset in the fit values of $f_{{\sigma}_v}$ in both figures~\ref{fig:nb_point_diff} and~\ref{fig:1st_par_diff}, but it remains at the same level as in section~\ref{sec:accuracy}, that is below the statistical uncertainty expected from the early DESI ELG sample.
 
\section{Conclusions}
\label{sec:conclusions}
In this paper, we introduced a method to fit HOD parameters using Gaussian Processes (GP) to provide a model of the multidimensional likelihood function, in the framework of a stochastic HOD modelling technique based on mock galaxy catalogues built from N-body simulations.

Our two-step procedure starts with initial training of the GP with 600 points distributed in the HOD parameter space according to Hammersley sampling. The likelihood model provided by the GP from this initial training is further improved by an iterative procedure adding one point to the training sample at each iteration, the next point to be added being randomly selected in Monte Carlo Markov chains (MCMC) run on the likelihood posterior predicted by the GP at the current iteration. This ensures that the sampling is made denser close the maximum of the likelihood function so as to provide a good determination of both the likelihood maximum and the error contours, despite the stochastic nature of our HOD modelling. The iterative procedure is pushed until a total of 800 iterations is achieved.

The reproducibility and accuracy of the method were studied on simulated mocks built from the \absum suite of high-accuracy N-body simulations on cubic boxes of 1~Gpc$/h$ length. These mocks are representative of the  expected density of the DESI ELG sample, but cover a volume three times larger than that covered by the early DESI ELG data. The procedure was repeated on sets of simulated mocks corresponding to different realisations of the same 6-parameter HOD model suitable for ELGs. Results on the 6 HOD parameters, defined by the marginalised values from the posterior distributions extracted from the MCMC chains run at the final iteration, were found to be reproducible within ranges smaller than those expected when cosmic variance is also included. In the presence of cosmic variance, we reach accuracies on the HOD parameters which are below the statistical uncertainty expected for early DESI ELG data, reaching at most 60$\%$ of the statistical uncertainty in the worst case (one parameter out of six, the maximum bias for the other parameters being 40$\%$).

We also explore the stability of the method when varying different ingredients. We find that the results do not depend on the sampling algorithm applied to define the training sample nor on the GP kernel. This is also true for the choice of the parameter with equidistant points in the initial training sample with our baseline Hammersley sampling. More dependence is found with respect to the choice of the number of points in the training sample and in the number of iterations after initial training. Different numbers of training points as well as numbers of iterations lower than 800 were tested. 
We find that there is no need to increase the density of the initial sampling infinitely. What matters most is to increase the density in the region of interest close to the likelihood maximum, once that region is roughly defined.  
In our framework this is achieved with our baseline of 600 points in initial training and 800 further iterations.

Finally, the fit progress towards stability during the iterative loop was monitored with the help of the Kullback-Leibler (KL) divergence between the MCMC chains. Requiring the KL divergence to be below 0.1 in a set of 20 consecutive iterations as a chain stability criterion, we observe that 96$\%$ of the fits pass this criterion well before iteration 800 and the few fits which fail are not outliers in any of the HOD parameters. On the other hand, if fits were stopped as soon as the KL criterion was met, we would obtain larger biases in the HOD parameters, showing that this does not ensure unbiased results. Hence our choice to push fits up to a total of 800 iterations, for which our tests on simulation show that the expected bias in the HOD parameter values is reasonably below the statistical uncertainty we expect from data. More generally, this illustrates the difficulty to define a robust convergence criterion when inference is performed on a surrogate model of a likelihood posterior while the model is still under evolution and subject to noise in the likelihood estimates.

Applying this procedure to DESI early data will be the subject of a forthcoming paper.

\bibliographystyle{JHEP}
\bibliography{references}{}

\begin{appendices}
\section{Practical implementation}
\label{app:app1}
The HOD fitting procedure described in this paper relies on an HOD pipeline and a fitting pipeline. Implementation details for both steps are given in the following.

\subsection{HOD pipeline}
The HOD pipeline\footnote{\url{ https://github.com/antoine-rocher/GP_HODpy}} built for this work produces mock catalogues and clustering measurements from an N-body simulation box, given parameters of an HOD model.
Besides the GHOD model defined in section~\ref{sec:hod}, the code supports four other HOD models for central galaxies, all supplemented by a power law model for satellite galaxies. By default, satellites are drawn within randomly pre-computed $r/r_s$ points from an NFW profile~\cite{NFW}. Satellites populated on simulation DM particles are also supported as an alternative. For each mock computation, central and satellite galaxies are generated independently and concatenated in a Python dictionary. To optimize the mock generation we use \texttt{numba}\footnote{\url{https://numba.pydata.org/}}, an open source Just In Time (JIT) compiler, that translates a subset of Python and \texttt{numpy} code into fast machine code, multi-threaded with automatic parallelization of JIT. We are also developing an MPI implementation of the mock generation that was not used for this study.  

Since HOD fitting results depend on the observed density of objects, the density in mocks can be set to a given value, as used in this work. The code generates the exact chosen density by pre-computing the total number of galaxies, $n_{gal}$, using eq. \ref{ngal} for the current set of HOD parameters. The amplitudes for centrals and satellites, $A_c$ and $A_s$, are then rescaled by $n_{gal,exp}/n_{gal}$, where $n_{gal,exp}$ is the galaxy number expected for the chosen density. As the resulting mock clustering only depends on the ratio ${A_s}/{A_c}$,  we can re-scale both amplitudes by the same factor to change the density while keeping the same clustering.

Once a mock catalogue is computed, the HOD pipeline provides an easy way to compute clustering measurements for the projected 2 point correlation function, $w_p$ and for the 2 point correlation function monopole and quadrupole, with user-defined separation ranges and binnings. 
We use the DESI wrapper \textsc{pycorr}\footnote{\url{https://github.com/cosmodesi/pycorr}} around the \texttt{Corrfunc} package~\cite{Corrfunc} 
to compute these measurements.

The fitting procedure requires $N=20$ mocks to be created at each point of the HOD parameter space to compute the $\chi^2$ value and its uncertainty.
To speed up the fitting procedure, 
this step runs the N mocks in parallel using the 
\texttt{joblib} package\footnote{\url{https://joblib.readthedocs.io/en/latest/generated/joblib.Parallel.html}}. Then, the N correlation functions are computed one after the other.
Reproducibility of the results when using multi-threading can be quite complicated. As each thread will generate a different seed (even if a seed is fixed at initialisation), it becomes difficult to have reproducible results. We choose to adopt a solution easy to implement.
We fix a single seed at initialisation and use it to determine a list of random integer numbers that will be used to initialise one seed per thread. This is easy to implement but the reproducibility of the results will depend on the number of threads.

\subsection{Fitting pipeline}
The Hammersley sampling of the HOD parameter space is performed with the PySMO sampling method\footnote{\url{https://idaes-pse.readthedocs.io/en/1.5.1/surrogate/pysmo/pysmo\_sampling\_properties.html}} of the \texttt{idaes-pse} package.
The Gaussian process part relies on the Gaussian Process Regressor\footnote{\url{https://scikit-learn.org/stable/modules/generated/sklearn.gaussian\_process.GaussianProcessRegressor.html}} function from the \texttt{scikit-learn} package.
The MCMC component is ensured by the \texttt{emcee} package and runs 12 chains of 10,000 points each in parallel, the first 800 points being discarded in each chain. 

\clearpage
\section{More on stability}
\label{app:conv}

\begin{figure}[tbp]
\centering
\includegraphics[width=\textwidth]{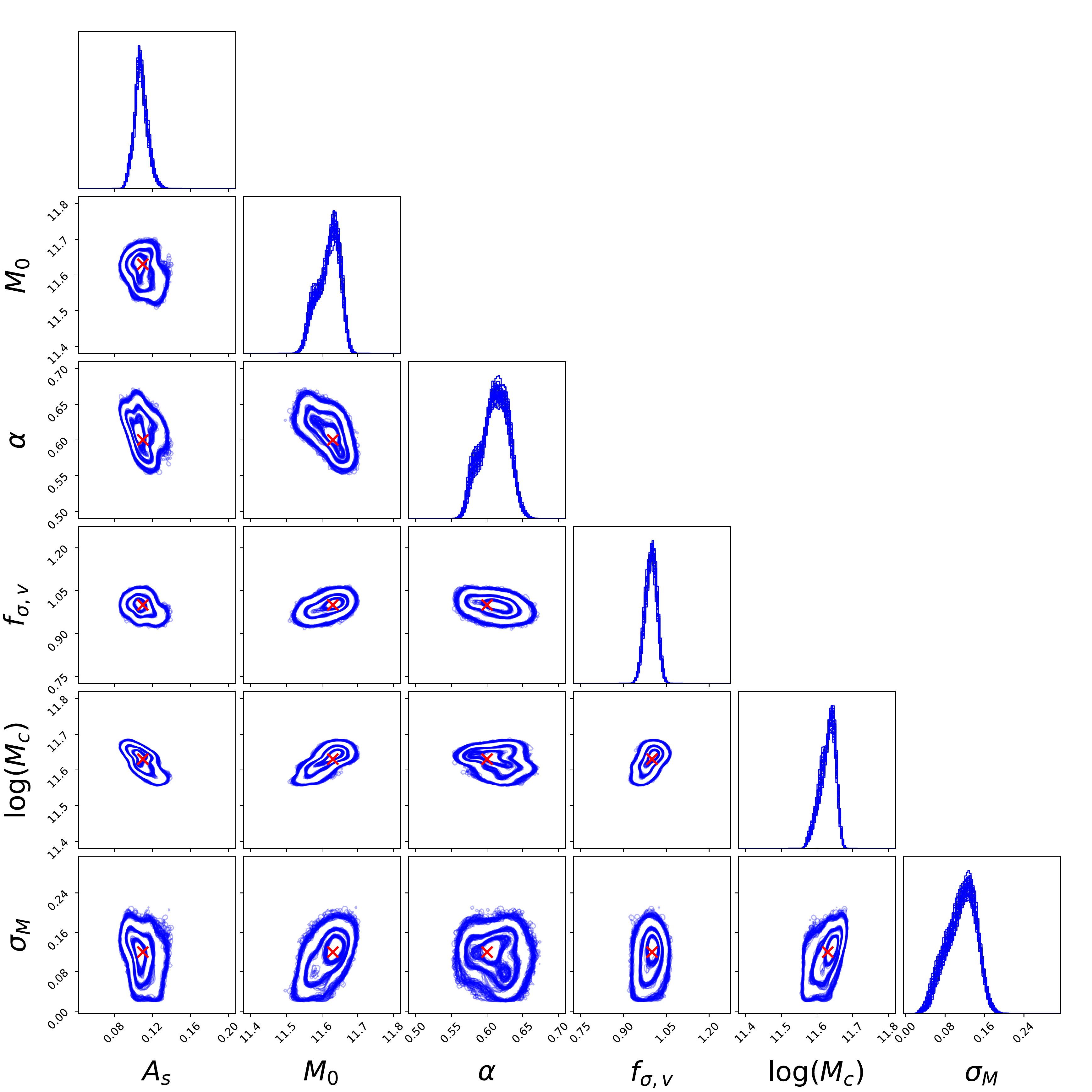} 
\caption{\label{fig:sup_contour} Contours and marginalised 1D-posteriors superimposed taking one iteration out of four between iterations 600 and 800 (so 50 iterations in total), for the fit shown in Figure~\ref{fig:convergence}.  The red cross is the parameter input values.
}
\end{figure}

Figure~\ref{fig:sup_contour} shows the superimposition of contours and marginalised 1D-posteriors from 50 iterations of the fit shown in Figures~\ref{fig:convergence} and~\ref{fig:ite800}. We took every one iteration out of four between iteration 600 and the stopping iteration, 800. This plot is complementary to Figure~\ref{fig:convergence}. It illustrates that, in the last quarter of iterations, the evolution of the GP surrogate model does not alter the marginalised median value of the posteriors but only changes slightly their $[16-84]$ percentile range.

\clearpage
\section{More on reproducibility fits}
\label{app:repro}

\begin{figure}[tbp]
\centering
\begin{tabular}{c}
\includegraphics[width=\textwidth]{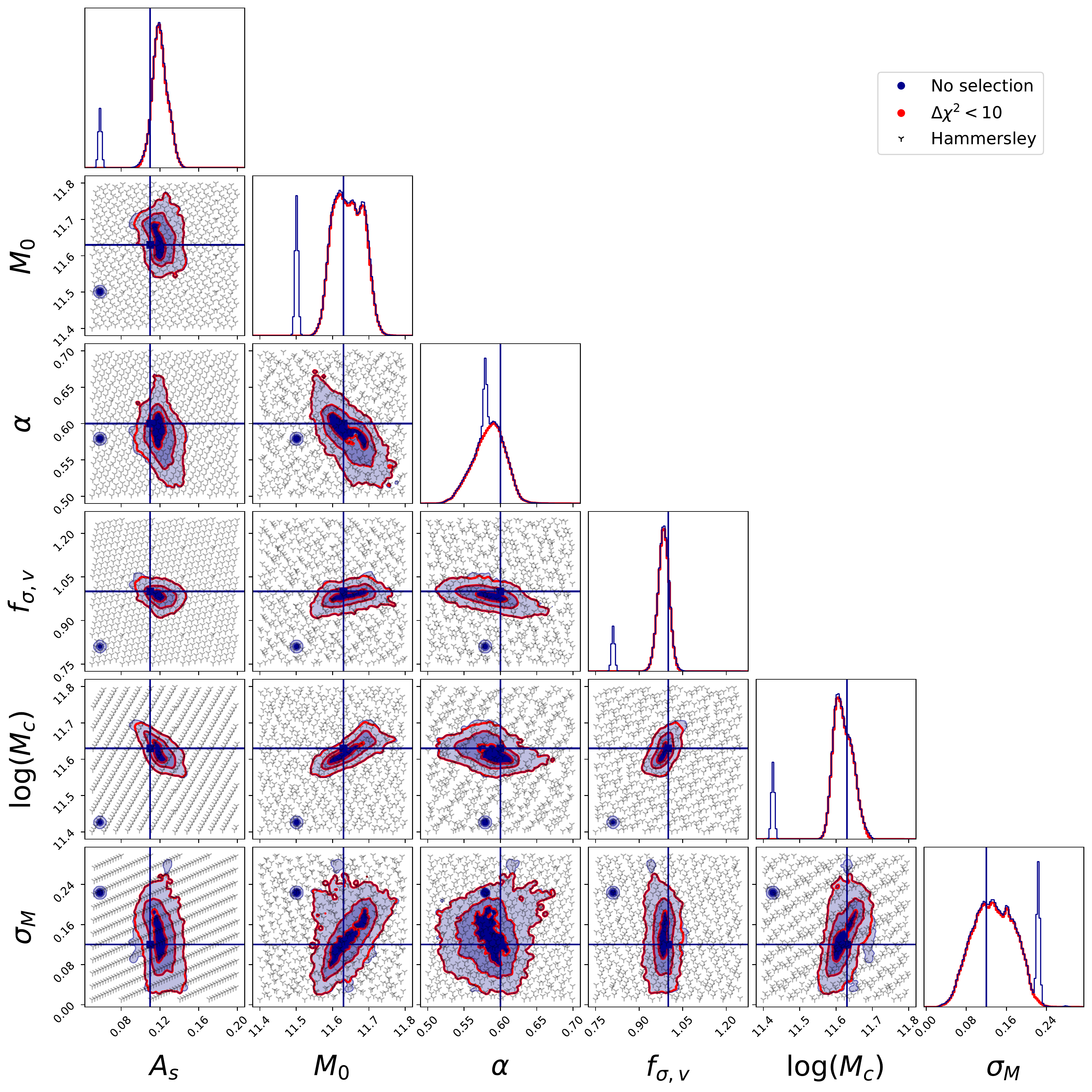} \\
\includegraphics[width=\textwidth]{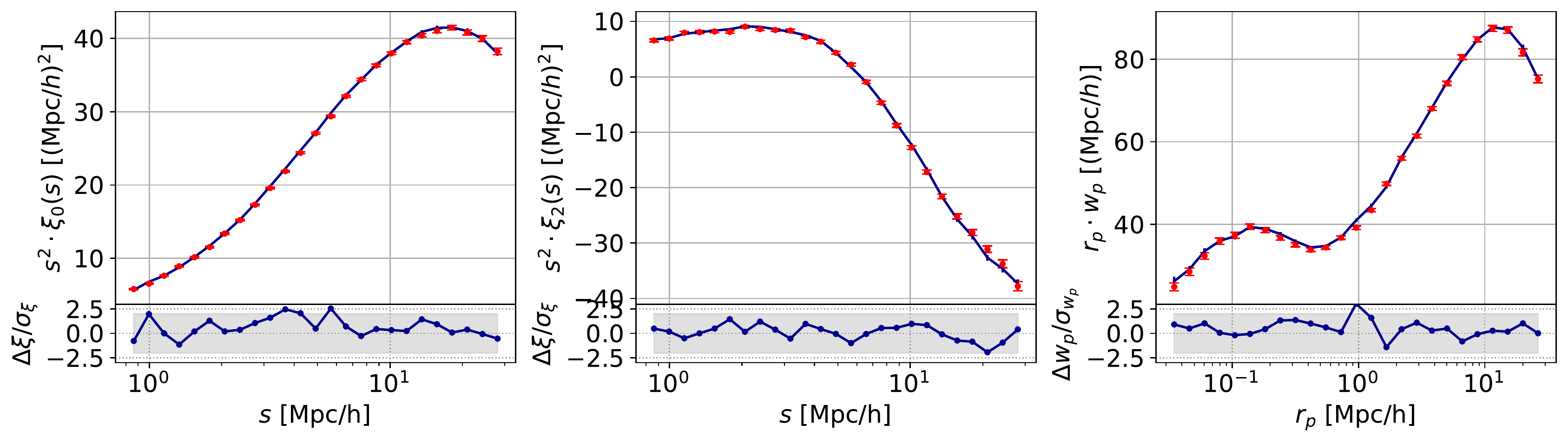}
\end{tabular}
\caption{\label{fig:cleaning} {\it Top.} Contours and marginalised 1D-posteriors  at iteration 800 for one fit from
 the reproducibility tests.  Solid lines are the input parameter values.
 Grey points are the Hammersley training sample. 
 The contours are obtained from MCMC chains after burning phase, with no further selection in blue and excluding 
 points with a large predicted $\chi^2$ error in red. {\it Bottom.} Clustering measurements predicted by the HOD model given by the marginalised values from the red 1D-posteriors in the top plot. The shaded band encompasses $\pm 2\sigma$ residuals, with errors on pseudo-data (red) and model (blue) added in quadrature.
}
\end{figure} 

In reproducibility fits, due to uncertainties on pseudo-data being very small, the surrogate model of the likelihood surface can present spikes, likely due to the stochasticity of the HOD modelling. Spikes occur in about 50$\%$ of the iterations and their locations in the HOD parameter space vary in the course of the iterative procedure. Besides, iterations showing spikes are generally associated with large $\chi^2$ errors in the MCMC chains, showing that these spikes are most probably spurious and mostly due to regions with not enough points, so the GP can predict huge variations in the $\chi^2$ value. In order to obtain reliable contours and errors, we remove points with large GP predicted $\chi^2$ errors in the MCMC chains. This is illustrated in Figure~\ref{fig:cleaning} which shows that spurious spikes do indeed disappear, leaving the main component of the contours unchanged. This procedure was applied to reproducibility fits only, all other fits have smooth predicted likelihood surface at the final iteration.

\end{appendices}

\end{document}